\documentclass[twocolumn%
,secnumarabic%
,amssymb,aps,prb,nobibnotes,longbibliography,superscriptaddress]{revtex4-1}
\usepackage{amsmath}
\usepackage{graphicx}
\usepackage{float}
\usepackage{color}
\usepackage{xr}

\AtBeginDocument{%
    \newwrite\bibnotes
    \def\bibnotesext{Notes.bib}
    \immediate\openout\bibnotes=\jobname\bibnotesext
    \immediate\write\bibnotes{@CONTROL{REVTEX41Control}}
    \immediate\write\bibnotes{@CONTROL{%
    apsrev41Control,author="08",editor="1",pages="1",title="0",year="1"}}
     \if@filesw
     \immediate\write\@auxout{\string\citation{apsrev41Control}}%
    \fi
}%

\expandafter\ifx\csname package@font\endcsname\relax\else
 \expandafter\expandafter
 \expandafter\usepackage
 \expandafter\expandafter
 \expandafter{\csname package@font\endcsname}%
\fi
\DeclareRobustCommand\substyle{\name@idx{document substyle}}%
\DeclareRobustCommand\classoption{\name@idx{document class option}}%
\DeclareRobustCommand\classname{\name@idx{document class}}%
\def\name@idx#1#2{%
 {\ttfamily#2}%
 \index{#2\space#1=\string\ttt{#2}\space#1}\index{#1>#2=\string\ttt{#2}}%
}%

\begin{document}
\title{Topological Edge Conduction Induced by Strong Anisotropic Exchange Interactions}
\author{Shehrin Sayed}\email{shehrinsayeed@gmail.com}
\affiliation{Materials Sciences Division, Lawrence Berkeley National Laboratory, Berkeley, CA 94720, USA.}
\affiliation{Electrical Engineering and Computer Sciences, University of California, Berkeley, CA 94720, USA.}
\author{Pratik Brahma}
\affiliation{Electrical Engineering and Computer Sciences, University of California, Berkeley, CA 94720, USA.}
\author{Cheng-Hsiang Hsu}
\affiliation{Electrical Engineering and Computer Sciences, University of California, Berkeley, CA 94720, USA.}
\author{Sayeef Salahuddin}\email{sayeef@berkeley.edu}
\affiliation{Materials Sciences Division, Lawrence Berkeley National Laboratory, Berkeley, CA 94720, USA.}
\affiliation{Electrical Engineering and Computer Sciences,
	University of California, Berkeley, CA 94720, USA.}

\begin{abstract}
We predict that an interplay between isotropic and anisotropic exchange interactions in a honeycomb lattice structure can lead to topological edge conduction when the anisotropic interaction is at least twice the strength of the isotropic interaction. For materials like Na$_2$IrO$_3$, such a strong anisotropic exchange interaction simultaneously induces a zigzag type of antiferromagnetic order that breaks the time-reversal symmetry of the topological edge conductor. We show that the electronic transport in such topological conductors will exhibit a quantized Hall conductance without any external magnetic field when the Fermi energy lies within a particular energy range.
\end{abstract}

\maketitle

\begin{figure}
	\includegraphics[width=0.48 \textwidth]{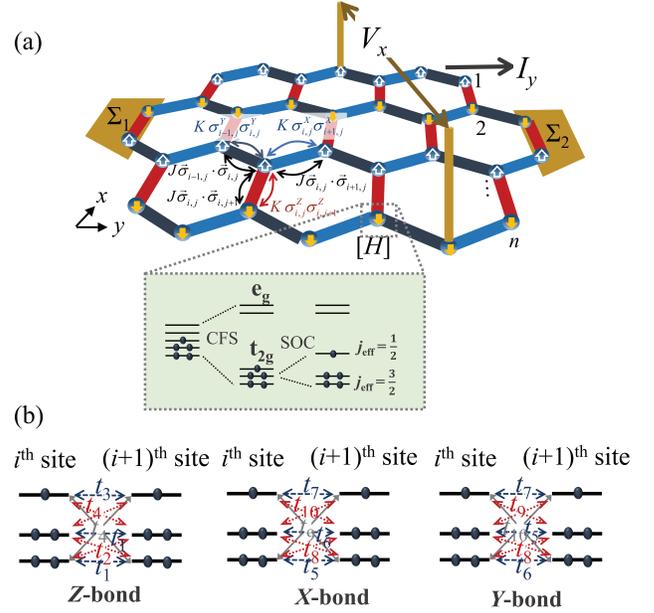}
	\centering
	\caption{(a) A honeycomb lattice with isotropic ($J$) and bond-dependent anisotropic ($K$) exchange interactions. Local $d$-orbital states on each lattice point in the presence of Coulomb interaction ($U$), crystal field splitting (CFS), and spin-orbit coupling (SOC). (b) General nearest neighbor hopping scenario along different bonds.}\label{fig0}
\end{figure}

Recently, there is a growing interest in transition metal-based oxides and halides with honeycomb lattice structure \cite{Savary_2016, Hermanns_AnnRev_2018, Winter_2017} (see Fig. \ref{fig0}(a)) for exhibiting both isotropic and bond-dependent anisotropic exchange interactions \cite{KITAEV20062}. An interplay between these exchange interactions can lead to quantum magnetism \cite{PhysRevLett.110.097204, PhysRevB.85.180403, Do2017, PhysRevB.83.220403, PhysRevB.91.144420, PhysRevB.92.235119, PhysRevB.86.224417} and a spin liquid state \cite{Banerjee1055, PhysRevB.93.235146, Takagi2019, Do2017}, which are modeled using the Kitaev-Heisenberg (KH) model as given by
\begin{equation}
\label{Full_KH_Model}
    \mathcal{H} = \sum {\,{K^\gamma }\,S_i^\gamma S_j^\gamma }  + \sum {J\,{{\vec S}_i} \cdot {{\vec S}_j}},
\end{equation}
where $\gamma\equiv \left\{x,y,z\right\}$, $K^\gamma$ is the anisotropic exchange interaction along $\gamma$ bond, $J$ is the isotropic exchange interaction, and $\vec S_i=\sum_{\gamma\equiv \left\{x,y,z\right\}}\hat{\gamma} \,S^\gamma_i$ is the spin-$\frac{1}{2}$ operator on the $i^\text{th}$ lattice point. 

In this letter, we predict that an interplay between the isotropic and anisotropic exchange interactions can form topological edge states when the anisotropic exchange interaction is at least twice the strength of the isotropic exchange interaction. Such an interplay will simultaneously induce an intrinsic magnetic order, specifically a zigzag antiferromagnetic (AFM) order, that breaks the time-reversal symmetry (TRS) of the edge conductor. We use a tight-binding model for such materials \cite{PhysRevB.93.214431, PhysRevB.90.155126, PhysRevLett.112.077204}, the KH model, and a non-equilibrium Green's function (NEGF)-based quantum-transport model \cite{Datta1995} to show that the exchange interaction induced topological edge conduction will exhibit a Hall conductance quantized to $q^2/h$ (where $q$ is the electron charge and $h$ is the Planck's constant) when the current is running along the zigzag direction and the voltage is measured along the armchair direction. We further calculate the band structures and Hall conductances of Na$_2$IrO$_3$ and $\alpha$-RuCl$_3$ and show that Na$_2$IrO$_3$ could be a model material to observe the phenomena predicted in this letter. Finally, using one-dimensional analytical arguments, we show that the topological edge state formation for high anisotropic exchange interaction is a property of Eq. \eqref{Full_KH_Model}. 

\begin{figure}
	%\vspace*{1cm}
	\includegraphics[width=0.5 \textwidth]{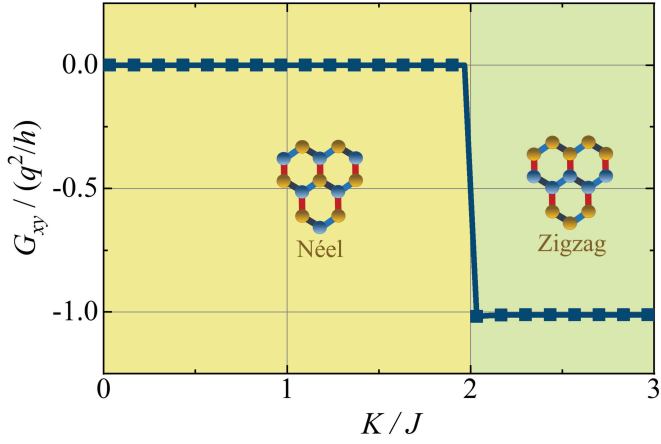}
	\centering
	\caption{The Hall conductance ($G_{xy}$) and corresponding magnetic order as a function of $K/J$.}\label{fig1}
\end{figure}

Transition-metal-based oxides and halides with honeycomb lattice structures have a half-filled $d^5$ ion (e.g., Ir$^{4+}$ or Ru$^{3+}$) in each of the lattice points in an octahedral environment. The crystal field (CF) splits the $d$-orbital into a $e_g$ (equally degenerate) pair and $t_{2g}$ (triply degenerate) states, see Fig. \ref{fig0}(a), and all five electrons occupy the $t_{2g}$ states. The presence of spin-orbit coupling (SOC) further splits the $t_{2g}$ states into $j_\text{eff}=\frac{1}{2}$ and $\frac{3}{2}$, states (see Fig. \ref{fig0}(a)), where the $j_\text{eff}=\frac{3}{2}$ states are filled and the $j_\text{eff}=\frac{1}{2}$ state is half-filled. Thus we have one hole per site. We describe the hole-mediated transport in such materials using a tight-binding Hamiltonian, given by
\begin{equation}
\label{sophis_H}
    \mathcal{H}=H_{U}+H_\text{CF}+H_\text{SOC}+H_{M}+H_\text{hop.},
\end{equation}
which is in the basis of the $d_{xy}$, $d_{yz}$, and $d_{zx}$ orbitals with each orbitals having two spin states, i.e., $\Psi\equiv \left\{\psi_{xy}^\uparrow,\psi_{xy}^\downarrow,\psi_{yz}^\uparrow,\psi_{yz}^\downarrow,\psi_{zx}^\uparrow,\psi_{zx}^\downarrow\right\}$. Eq. \eqref{sophis_H} takes into account the static inter and intra -orbital Coulomb repulsion ($H_{U}$), crystal field ($H_\text{CF}$), spin-orbit coupling ($H_\text{SOC}$), local fields at lattice sites generated by magnetic ordering ($H_M$) according to the solution of Eq. \eqref{Full_KH_Model}, and hole-based hopping ($H_\text{hop.}$) with bond-dependent hopping integrals ($t_{1}$ to $t_{10}$) shown in Fig. \ref{fig0}(b). The effective values of the hopping integrals are adopted from Ref. \cite{PhysRevB.93.214431}, which were obtained by fitting detailed DFT calculations considering electron-electron interactions.

The isotropic ($J$) and anisotropic ($K^z=K^x=K^y=K$) exchange interaction strengths in Eq. \eqref{Full_KH_Model} are related to these hopping integrals as \cite{PhysRevB.93.214431, PhysRevB.90.155126, PhysRevLett.112.077204}
\begin{subequations}
\begin{equation}
    J = \delta \, {\left( {2{t_1} + {t_3}} \right)^2} - \xi \left\{ {9t_4^2 + 2{{\left( {{t_1} - {t_3}} \right)}^2}} \right\},
\end{equation}
\begin{equation}
    K = \xi \left\{ {3t_4^2 + {{\left( {{t_1} - {t_3}} \right)}^2} - 3t_2^2} \right\}.
\end{equation}
\end{subequations}
where the coefficients $\delta$ and $\xi$ are determined by the Coulomb repulsion, Hund's coupling, and SOC strengths \cite{PhysRevB.93.214431, PhysRevB.90.155126, PhysRevLett.112.077204}. We have solved Eq. \eqref{Full_KH_Model} in Fock space to find the ground energy state and corresponding intrinsic magnetic order for a particular combination of $J$ and $K$ calculated using the tight-binding parameters for Na$_2$IrO$_3$ \cite{PhysRevB.93.214431}. %Here, $\delta$ and $\xi$ are 1.176 eV$^{-1}$ and 0.098 eV$^{-1}$, respectively.

\begin{figure}
	%\vspace*{1cm}
	\includegraphics[width=0.5 \textwidth]{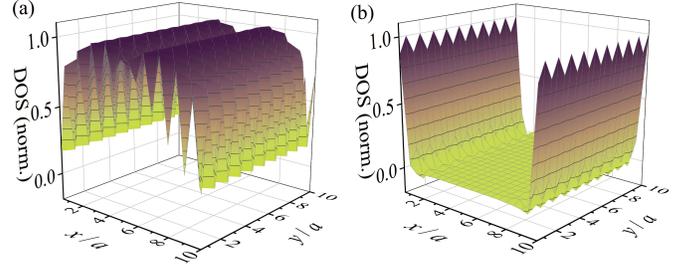}
	\centering
	\caption{Local density of states on the two-dimensional honeycomb lattice structure in Fig. \ref{fig0}(a) for (a) $K=0.1J$ and (b) $K=2.5J$. }\label{fig2}
\end{figure}

We use the tight-binding model in Eq. \eqref{sophis_H} to calculate the Green's function $\mathcal{G}=\left[E\mathcal{I}-\mathcal{H}-\Sigma_0-\Sigma_1-\Sigma_2\right]^{-1}$ for the structure in Fig. \ref{fig0}(a), where $E$ is the energy, $\mathcal{I}$ is the identity matrix, $\Sigma_{1,2}$ are the self-energy functions of the left and right contacts, respectively, $\Sigma_0 = \mathcal{W}\, \mathcal{G}$ is the self-energy that takes into account the dephasing in the channel due to electron-electron interactions in a self-consistent manner. Here $\mathcal{W}_{m,n}=\langle\,U_m\,U_n^*\rangle$ is calculated from the random potential at $m^\text{th}$ and $n^\text{th}$ lattices. We use Green's function \cite{Datta1995} to calculate the potential distribution on the $xy$-plane as
\begin{equation}
    \label{pot_distribution}
    V(x,y) =V_0\; \mathrm{Re}\left\{\dfrac{\mathcal{G} \left(\Gamma_1 f_1+\Gamma_2 f_2\right) \mathcal{G}^\dagger}{\mathcal{G}\,\left(\Gamma_1+\Gamma_2 \right) \mathcal{G}^\dagger}\right\},
\end{equation}
where $\Gamma_{1,2}=j\left(\Sigma_{1,2}-\Sigma_{1,2}^\dag\right)$ are the broadening functions and represent the anti-Hermitian parts of the self-energy functions, $f_{1,2}$ are the Fermi occupation factors of the left and right contacts respectively, and $V_0$ is the applied potential. The Hall resistance is given by
\begin{equation}
\label{Hall_Resistance}
    R_{xy} = \dfrac{\Delta V_x}{I_y} = \dfrac{V(x=0,y)-V(x=w,y)}{\dfrac{q}{h}\int dE\,\,\mathrm{Trace}\left[\mathrm{Re}\left\{\Gamma_1 \mathcal{G} \Gamma_2\mathcal{G}^\dag \right\}\right] \, \left(f_1-f_2\right)},
\end{equation}
where the numerator represents the voltage difference between the two edges along $x$-direction and the denominator represents the current $I_y$ flowing along the $y$-direction. Here, $w$ is the channel width. The Hall conductance $G_{xy}$ is calculated using $R_{xy}$ and longitudinal resistance $R_{xx}$ as $G_{xy} = {R_{xy}}/({R_{xy}^2+R_{xx}^2})$.

In order to understand the effect of the intrinsic exchange interactions in the transport properties, we calculate $G_{xy}$ and the magnetic order of the honeycomb lattice as a function of the relative strength between the anisotropic and isotropic exchange interactions (i.e., $|K/J|$), as shown in Fig. \ref{fig1}. We apply a current $I_y$ along the zigzag chains ($y$-direction) and calculate the transverse voltage $V_x$ along the armchair direction ($x$-direction), see Fig. \ref{fig0}(a). We change the $|K|/|J|$ ratio by changing the hopping integral $t_4$ and keep other parameters equivalent to the parameter values for Na$_2$IrO$_3$. The NEGF calculations show that $|G_{xy}|$ is 0 when $K < 2J$; however,  becomes exactly $q^2/h$ when the $K \geq 2J$, see Fig. \ref{fig1}. This phenomenon induced by a strong anisotropic exchange interaction in a two-dimensional (2D) channel is very similar to the popular signature of a quantum anomalous Hall (QAH) state \cite{Chang167,PhysRevLett.120.056801,Mogieaao1669,Deng895} observed in band-inverted three-dimensional magnetic topological insulators with broken time-reversal symmetry (TRS).

The quantized Hall conductance in Fig. \ref{fig1} is a combination of two effects: topological edge state formation when $K \geq 2J$ and intrinsic magnetic ordering induced by the interplay between $J$ and $K$ that simultaneously breaks the TRS. In order to understand the edge state formation, we calculate the local density of states as 
\begin{equation}
    \mathcal{D} = \frac{1}{{2\pi }}\text{Re}\left( \mathcal{G}\,\Gamma\, \mathcal{G}^\dagger \right),
\end{equation}
where $\Gamma = j\left\{\Sigma_1+\Sigma_2-\left(\Sigma_1+\Sigma_2\right)^\dagger\right\}$ is the broadening function. The spatial distribution of the local density of states is shown in Fig. \ref{fig2}(a) and (b) for tight-binding parameters that correspond to $K<2J$ and $K>2J$, respectively. The density of states is non-zero everywhere throughout the 2D channel when $K<2J$ and there are no distinct edge states, as shown in Fig. \ref{fig2}(a) for the case $|K|/|J| \approx 0.1$. For $K>2J$, the density of states in the middle of the 2D channel becomes zero, i.e., the middle of the 2D channel becomes insulating. However, the case of $K>2J$ exhibits edge states as shown in Fig. \ref{fig2}(b) for $|K|/|J| \approx 2.5$.

The honeycomb lattice considered here has a N\'eel type AFM order for $K < 2J$. Thus, the two-dimensional channel in the honeycomb lattice structure does not have an effective field that can produce an anomalous Hall effect and does not contribute to $G_{xy}$. Thus, we observe $G_{xy}=0$. However, there is a sharp transition from N\'eel type AFM order to a zigzag type AFM order for $K > 2J$, see Fig. \ref{fig1}. A zigzag AFM order refers to the case where each of the zigzag atomic chains is a one-dimensional (1D) ferromagnetic chain; however, the ferromagnetic order alternates along the armchair direction (see Fig. \ref{fig0}(a)), giving rise to an overall AFM order. Thus, the edge atomic chains have a net magnetic order that breaks the TRS of the edge conductor. Note that the magnetic order for the case $K=2J$ is a superposition of a N\'eel and a zigzag AFM states; see supplementary information for details.

\begin{figure}
	%\vspace*{1cm}
	\includegraphics[width=0.5 \textwidth]{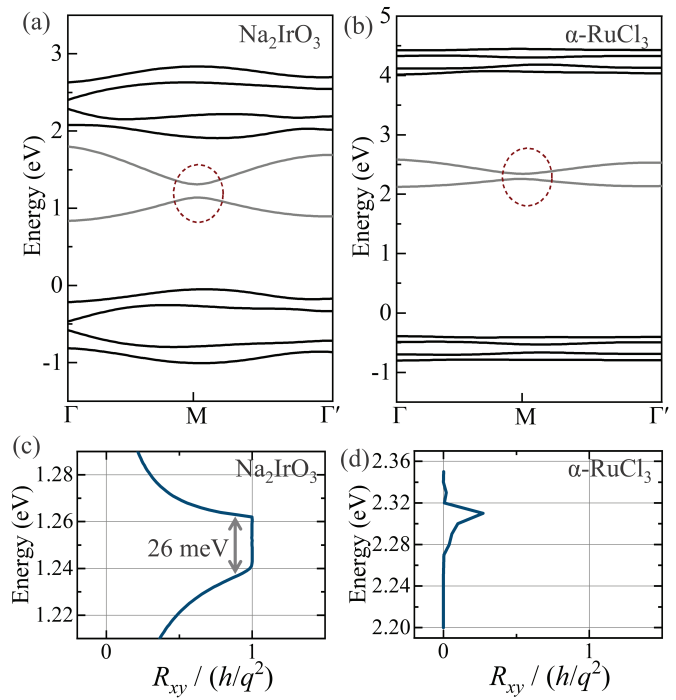}
	\centering
	\caption{Band diagram for (a) Na$_2$IrO$_3$ and (b) $\alpha$-RuCl$_3$. Hall resistance is (c) quantized to $h/q^2$ for Na$_2$IrO$_3$ within the gap at $M$ symmetry point. (d) No quantized Hall effect is observed for $\alpha$-RuCl$_3$. }\label{fig3}
\end{figure}

The zigzag AFM order has been observed in the transition metal oxides and halides, e.g., Na$_2$IrO$_3$ and $\alpha$-RuCl$_3$, at low temperatures \cite{PhysRevLett.110.097204, PhysRevB.85.180403, Do2017, PhysRevB.83.220403, PhysRevB.92.235119, PhysRevB.91.144420}. We calculate the band diagram of Na$_2$IrO$_3$ and $\alpha$-RuCl$_3$ using Eq. \eqref{sophis_H}, see Fig. \ref{fig3}(a)-(b), where the tight-binding parameters correspond to a zigzag AFM order. We observe the spin bands for the $j_\text{eff}=\frac{1}{2}$ energy states show a small gap of $\sim26$ meV near the $M$ symmetry point for both materials. We have calculated $R_{xy}$ using Eq. \eqref{Hall_Resistance} as a function of energy within the energy gap and found it to be quantized to $h/q^2$ for Na$_2$IrO$_3$, as shown in Fig. \ref{fig3}(c). This indicates that the energy gap hosts non-trivial topological edge states. Interestingly, the calculated $J\approx3.4$ meV and $K\approx-22.3$ meV, thus yields $|K|/|J|\approx 6.6$ and satisfies the condition in Fig. \ref{fig1}. This phenomenon will be observed in Hall measurements on Na$_2$IrO$_3$ at low temperatures where zigzag AFM forms and as long as the Fermi energy lies within the topological energy gap.

For $\alpha$-RuCl$_3$, we do not observe a quantized $R_{xy}$ in the transport calculation within the energy gap, see Fig. \ref{fig3}(d). This is consistent with the condition in Fig. \ref{fig1}, as $J$ and $K$ are $-2.02$ meV and $-3.4$ meV based on the tight-binding parameters, respectively, which gives $|K|/|J|\sim 1.68$. A non-zero $R_{xy}$ for a small energy range within the upper band is observed which corresponds to the anomalous Hall effect due to an uncompensated net magnetization in the lattice with a zigzag AFM order. Note that a half-integer quantized Hall effect has been observed for thermal transport \cite{Kasahara2018, Yokoi568} in $\alpha$-RuCl$_3$, which is out of the scope of this letter. In this letter, we predict an anisotropic exchange interaction induced quantized Hall conductance due to charge transport in the topological edge state and we do not expect any topological conduction in $\alpha$-RuCl$_3$.

Na$_2$IrO$_3$ has been predicted to exhibit a quantum spin Hall (QSH) state at the $M$ symmetry point \cite{PhysRevLett.102.256403}, which is equivalent to a 2D topological insulator with TRS \cite{Konig766} that exhibits a spin-momentum locked linear dispersion. Such a QSH state has not been observed yet \cite{PhysRevB.93.245132, PhysRevB.91.041405}; however, it has been pointed out that Na$_2$IrO$_3$ exhibits a persistent energy gap \cite{PhysRevB.91.041405}. Here, we point out that the energy band in Fig. \ref{fig3}(a) exhibits a small gap at the $M$ symmetry point and an intuitive explanation is that the TRS protected linear dispersion predicted in Ref. \cite{PhysRevLett.102.256403} should exhibit a gap induced by TRS breaking with a net intrinsic magnetic order in the material. Note that this gap is similar to that observed in magnetic topological insulator \cite{Chang167,PhysRevLett.120.056801,Mogieaao1669, Liu2020, Deng895}; however, such a gap in Na$_2$IrO$_3$ will be induced by the interplay between the isotropic and anisotropic exchange interactions.

Topological phases and QAH state have been theoretically discussed in various ferromagnetic oxides \cite{PhysRevLett.119.026402,PhysRevB.95.125430,PhysRevB.96.205433}, and halides \cite{PhysRevB.95.201402,PhysRevB.95.045113} due to strong SOC. In this letter, we identify a new mechanism for topological edge conduction, which can be observed in specific oxides and halides known for exhibiting strong anisotropic exchange interactions \cite{Savary_2016, Winter_2017, PhysRevB.93.214431}. We specifically discuss the Hall effect in the 2D channels with zigzag AFM order that will exhibit signatures similar to QAH.

We now show that the topological edge state formation for $K\geq2J$ is a property of the KH model in Eq. \eqref{Full_KH_Model}. For analytical simplicity, we start with a single atomic chain with $N$ lattice points by retaining the zigzag nature of the anisotropic interaction \cite{Agrapidis2018}, as
\begin{equation}
\label{KH_Model_1D}
\begin{aligned}
    \mathcal{H} = \sum\limits_m \left(J {{{\vec S}_{m + 1}} \cdot {{\vec S}_m}}  + {K^x} {S_{2m}^xS_{2m - 1}^x} + {K^y} {S_{2m + 1}^yS_{2m}^y}\right).
\end{aligned}
\end{equation}
%Similar one-dimensional (1D) model has been used in the past \cite{Agrapidis2018} to study phase diagram and magnetic structure under the approximation of a zigzag type ordering.

We represent the $S=\frac{1}{2}$ spin chains in Eq. \eqref{KH_Model_1D} using fermionic operators, $\left| \uparrow \right\rangle\equiv f^\dagger \left| \downarrow \right\rangle$ and $\left| \downarrow \right\rangle\equiv f \left| \uparrow \right\rangle$ using the Jordan-Wigner transformation \cite{PhysRevB.80.125415, PhysRevB.101.094403}, to get
\begin{equation}
\label{KH_1D_JW}
    \begin{array}{l}
\mathcal{H} = \dfrac{1}{2}\left( {J + \dfrac{{K}}{2}} \right)\sum\limits_m {\left( {f_{m + 1}^\dag {f_m} + f_m^\dag {f_{m + 1}}} \right)}- 2J\sum\limits_m {{n_m}} \\
 - {{\dfrac{{K}}{4}} \sum\limits_m {{{\left( { - 1} \right)}^m}\left( {f_{m + 1}^\dag f_m^\dag  + {f_m}{f_{m + 1}}} \right)}} + J\sum\limits_m {{n_{m + 1}}{n_m}},
\end{array}
\end{equation}
where a constant term $N/4$ is ignored. Here, $n_m=f_m^\dag f_m$ and we assume $K^x=K^y=K$.

We use Fourier transformation around the band minimum, $f_m = \frac{1}{{\sqrt N }}\sum_{k'} {{s_{k'}}{e^{i{{k'}}{m}}}}$ to convert Eq. \eqref{KH_1D_JW} from the real-space to the momentum space, as given by
\begin{equation}
\label{KH_1D_kspace}
\begin{aligned}
\mathcal{H} = \sum\limits_{k'} {\left\{ { 2J - {\frac{{2J + {K}}}{2}} \cos k'} \right\}s_{k'}^\dag {s_{k'}}}  \\+ \sum\limits_{k'} {\frac{{{K}}}{4}\sin k'\left( {s_{k'}}{s_{ - k'}} - {s_{ - k'}^\dag s_{k'}^\dag } \right)} \\+ \frac{1}{N}\sum\limits_{k',q} {J {{e^{-iqa}}} s_{k'-q}^\dag {s_{k'}^\dag} {s_{k'-q}}}s_{k'},
\end{aligned}
\end{equation}
where $k'=k-k_0$. Here, $k, q$ are the wavevector, $k_0$ is the wavevector at the band minimum, and $s_{k'}^\dagger$ is the creation operator in momentum-space. The third term in Eq. \eqref{KH_1D_kspace} represent the interaction term. Here, we consider a case where the atomic chain is a one-dimensional ferromagnet, corresponding to the case for the edge atomic chain for a zigzag AFM order. Under such approximation, the interaction term can be ignored for mathematical simplicity (see, e.g., Ref. \cite{coleman_2015}) to analyze a first-order dispersion relation. 

We transform Eq. \eqref{KH_1D_kspace} into a $2\times2$ matrix in the particle-hole basis using $\mathcal{H} = \sum_k \Psi^\dag\,\mathrm{H} \Psi$ with $\Psi^\dag \equiv \left\{s_{k'}^\dag,s_{-k'}\right\}$, which yields
\begin{equation}
\label{TB_KH_matrix}
\begin{array}{cc}
    \mathrm{H} = - J{\sigma _z} + \dfrac{{2J + K}}{4}{\sigma _z}\cos k' + \dfrac{{K}}{4}{\sigma _x}\sin k'.
\end{array}
\end{equation}
Eq. \eqref{TB_KH_matrix} is in the basis of particle-hole; however, here particle represents an up spin, and the hole represents a down spin. Thus, Eq. \eqref{TB_KH_matrix} has the basis of up and down spins. The energy gap near the band minimum (i.e., $k\rightarrow k_0$) is given by $|K-2J|/2$ (see supplementary information). The energy gap near the $M$ symmetry point in Fig. \ref{fig3}(a) for the Na$_2$IrO$_3$ case is $\sim$26 meV, which is very close to the analytical estimation $|K-2J|/2\approx$ 29 meV. Note that the $q^2/h$ plateau is observed within this energy gap (see Fig. \ref{fig3}(c)) since $K$ and $J$ satisfies the condition $K\geq 2J$.

We calculate the topological winding number \cite{YakovenkoPRB, PhysRevB.83.125109} by obtaining the eigenfunction, $\psi(k)$ of Eq. \eqref{TB_KH_matrix}, as
\begin{equation}
\label{winding_numb}
\begin{aligned}
    {W_Z} = \frac{1}{{j\pi }}\int_0^{2\pi } d k\;\psi {(k)^\dag }{\partial _k}\psi (k) = \left\{ \begin{array}{l}
1;\;\;{\rm{when}}\;K \ge 2J,\\
0;\;\;{\rm{ when}}\;K < 2J,
\end{array} \right.
\end{aligned}
\end{equation}
Here, $W_Z=0$ for $K<2J$, indicating a trivial state, while $W_Z=1$ for $K\geq2J$, indicating topological edge states occurring for a strong anisotropic exchange interaction. Similar one-dimensional arguments for surface state formation in two and three-dimensional topological insulators have been discussed in the past \cite{YakovenkoPRB}. The analytical conditions obtained from one-dimensional arguments in Eq. \eqref{winding_numb} exactly correspond to the full tight-binding model-based NEGF results in Fig. \ref{fig1}. Note that the full NEGF calculations consider a general case for the interactions.

In summary, we predict a strong anisotropic exchange interaction induced topological edge conduction in materials with honeycomb lattice structures. The topological edge conductor forms when the strength of the anisotropic interaction exceeds at least twice the strength of the isotropic interaction. For materials like Na$_2$IrO$_3$, such strong anisotropic exchange interaction simultaneously induces an intrinsic zigzag antiferromagnetic order that breaks the time-reversal symmetry of the edge conductor. We use a nearest-neighbor tight-binding model and the Kitaev-Heisenberg model to study the quantum transport in such materials and show that the existence of the time-reversal symmetry broken topological edge conductor will be exhibited as a quantized Hall conductance without any external magnetic field.

\begin{acknowledgments}
This work is in part by the U.S. Department of Energy, under Contract No. DE-AC02-05-CH11231 within
the NEMM program (MSMAG) and in part supported by ASCENT, one of six centers in JUMP, an SRC program sponsored by DARPA.
\end{acknowledgments}

\section*{Supplementary Information}
\appendix
\section{Transport modeling for materials with anisotropic exchange interactions}
\label{AppTB}

In this section, we will discuss the details of the tight-binding model used in the main manuscript.

\subsection{Tight-Binding Parameters}

\subsubsection{Coulomb Interaction Term}
We approximate the Coulomb term using a one-body operator under a one hole-based transport scenario as given by
\begin{equation}
\label{coulombH_mat}
    {H_U} = \left[ {\begin{array}{*{20}{c}}
U & U' & U'\\
U' & U & U'\\
U' & U' & U\\
\end{array}} \right]\otimes {\mathcal{I}_{2 \times 2}},
\end{equation}
where $U$ is the intraorbital coulomb repulsion and $U'=U-3J_H$ is the interorbital coulomb repulsion. The parameters are summarized in Table \ref{tab1}.

In addition, we consider the dephasing of the hole transport due to the Coulomb interaction with other electrons using a self-consistent approach using an additional self-energy function within the non-equilibrium Green's function method described in Appendix \ref{AppNEGF}.

\subsubsection{Crystal Field}

The crystal field (CF) is described by 
\begin{equation}
    {H_{CF}} = \left[ {\begin{array}{*{20}{c}}
0&{{\gamma _1}}&{{\gamma _2}}\\
{{\gamma _1}}&0&{{\gamma _2}}\\
{{\gamma _2}}&{{\gamma _2}}&{{\gamma _3}}
\end{array}} \right] \otimes {\mathcal{I}_{2 \times 2}},
\end{equation}
where $\gamma_{1,2}$ are crystal field parameters given in Table \ref{tab1} and $I_{2\times2}$ is a 2 by 2 identity matrix.

\subsubsection{Spin-Orbit Coupling}

The spin-orbit coupling (SOC) is described by
\begin{equation}
    {H_{SOC}} = \frac{{{\lambda _{SOC}}}}{2}\left[ {\begin{array}{*{20}{c}}
0&{ - j{\sigma _z}}&{j{\sigma _y}}\\
{j{\sigma _z}}&0&{ - j{\sigma _x}}\\
{ - j{\sigma _y}}&{j{\sigma _x}}&0
\end{array}} \right],
\end{equation}
where $\lambda_{SOC}$ is the spin-orbit parameter given in Table \ref{tab1} and $\sigma_{x,y,z}$ are the Pauli matrices.

\subsubsection{Hopping Integrals}
The hopping matrix along the $Z$-bond for the nearest-neighbor interaction is given by
\begin{equation}
    {T_Z} = \left[ {\begin{array}{*{20}{c}}
{{t_1}}&{{t_2}}&{{t_4}}\\
{{t_2}}&{{t_1}}&{{t_4}}\\
{{t_4}}&{{t_4}}&{{t_3}}
\end{array}} \right] \otimes {\mathcal{I}_{2 \times 2}},
\end{equation}
where $I_{2\times2}$ is a 2 by 2 identity matrix.

The hopping matrix along the $X$-bond he nearest-neighbor interaction is given by
\begin{equation}
{T_X} = \left[ {\begin{array}{*{20}{c}}
{{{t}_5}}&{{{t}_{8}}}&{{{t}_{9}}}\\
{{{t}_{8}}}&{{{t}_{6}}}&{{{t}_{10}}}\\
{{{t}_{9}}}&{{{t}_{10}}}&{{{t}_{7}}}
\end{array}} \right] \otimes {\mathcal{I}_{2 \times 2}}.
\end{equation}

The hopping matrix along the $Y$-bond he nearest-neighbor interaction is given by
\begin{equation}
{T_Y} = \left[ {\begin{array}{*{20}{c}}
{{{t}_6}}&{{{t}_{8}}}&{{{t}_{10}}}\\
{{{t}_{8}}}&{{{t}_{5}}}&{{{t}_9}}\\
{{{t}_{10}}}&{{{t}_9}}&{{{t}_{7}}}
\end{array}} \right] \otimes {\mathcal{I}_{2 \times 2}}.
\end{equation}
The parameter values used for calculation are summarized in Table \ref{tab1}.

\subsubsection{Magnetic Order}

The magnetic order of each lattice point for a given $J$ and $K$ combination is obtained from the Kitaev-Heisenberg model in Eq. \eqref{Full_KH_Model} in the main manuscript. The magnetic order term in the Hamiltonian is given by
\begin{equation}
    H_M = b \vec{S}_i\cdot \vec\sigma
\end{equation}

\begin{figure}
	%\vspace*{1cm}
	\includegraphics[width=0.45 \textwidth]{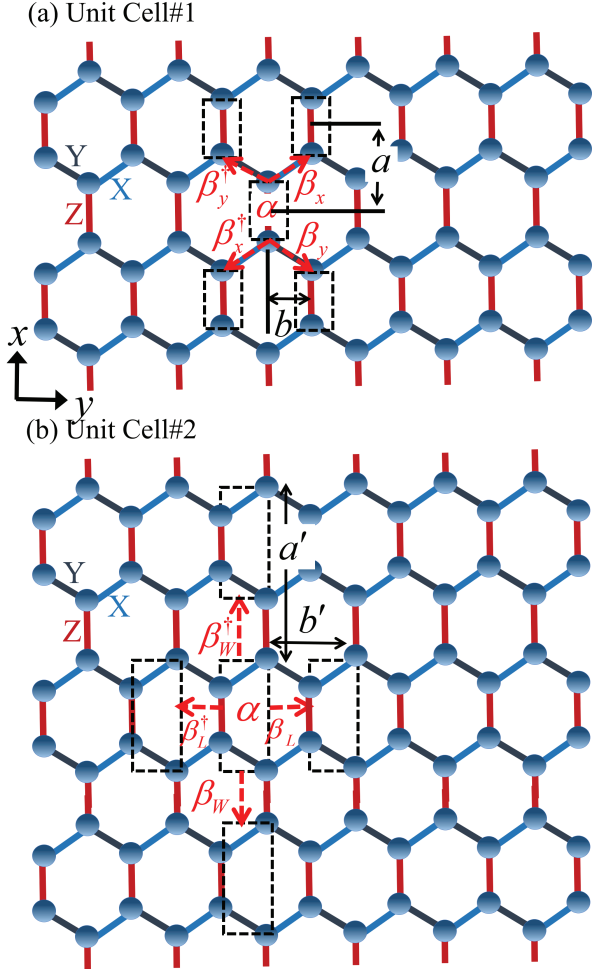}
	\centering
	\caption{The choice of unit cells for the tight-binding calculations in honeycomb lattice.}\label{fig7}
\end{figure}

\subsection{Tight-Binding Hamiltonian}
At a lattice point, the self energy matrix is given by
\begin{equation}
    \alpha_0 = H_U + H_{CF} + H_{SOC}.
\end{equation}
We have used two definitions of unit cell as shown in Fig. \ref{fig7} to form the corresponding tight-binding Hamiltonian. For the definition in Fig. \ref{fig7}(a), the form of the tight-binding Hamiltonian is given by
\begin{subequations}
\begin{equation}
    \begin{array}{l}
\mathcal{H}= \left[ \alpha  \right] + \left[ {{\beta _x}} \right]{e^{j\left( {{k_x}a + {k_y}b} \right)}} + \left[ {\beta _x^\dag } \right]{e^{ - j\left( {{k_x}a + {k_y}b} \right)}}\\
\,\,\,\,\,\,\,\,\,\,\,\,\,\,\, + \left[ {{\beta _y}} \right]{e^{-j\left( {{k_x}a - {k_y}b} \right)}} + \left[ {\beta _y^\dag } \right]{e^{ j\left( {{k_x}a - {k_y}b} \right)}},
\end{array}
\end{equation}
\begin{equation}
    \left[ \alpha  \right] = \left[ {\begin{array}{*{20}{c}}
\alpha_0&{{T_Z}}\\
{{T_Z}}&\alpha_0
\end{array}} \right],
\end{equation}
\begin{equation}
    \left[ {{\beta _x}} \right] = \left[ {\begin{array}{*{20}{c}}
0&{{T_X}}\\
0&0
\end{array}} \right],\;\;\;\;\text{and}
\end{equation}
\begin{equation}
    \left[ {{\beta _y}} \right] = \left[ {\begin{array}{*{20}{c}}
0&{{T_Y}}\\
0&0
\end{array}} \right].
\end{equation}
\end{subequations}
where $a$ and $b$ are the lattice constant along the $x$ and the $y$ directions, respectively.

For the definition in Fig. \ref{fig7}(b), the form of the tight-binding Hamiltonian is
\begin{subequations}
\begin{equation}
    \begin{array}{l}
\mathcal{H}= \left[ \alpha  \right] + \left[ {{\beta _W}} \right]{e^{j {{k_x}a'}} } + \left[ {\beta _W^\dag } \right]{e^{ - j {{k_x}a' }}}\\
\,\,\,\,\,\,\,\,\,\,\,\,\,\,\, + \left[ {{\beta _L}} \right]{e^{j { {k_y}b}}} + \left[ {\beta _L^\dag } \right]{e^{ -j {{k_y}b} }},
\end{array}
\end{equation}
\begin{equation}
    \left[ \alpha  \right] = \left[ {\begin{array}{*{20}{c}}
{{\alpha _0}}&{{T_Y}}&0&0\\
{T_Y^\dag }&{{\alpha _0}}&{{T_Z}}&0\\
0&{T_Z^\dag }&{{\alpha _0}}&{{T_X}}\\
0&0&{T_X^\dag }&{{\alpha _0}}
\end{array}} \right],
\end{equation}
\begin{equation}
    \left[ {{\beta _W}} \right] = \left[ {\begin{array}{*{20}{c}}
0&0&0&0\\
0&0&0&0\\
0&0&0&0\\
{{T_Z}}&0&0&0
\end{array}} \right],\;\;\;\;\text{and}
\end{equation}
\begin{equation}
    \left[ {{\beta _L}} \right] = \left[ {\begin{array}{*{20}{c}}
0&{{T_X}}&0&0\\
0&0&0&0\\
0&0&0&0\\
0&0&{{T_Y}}&0
\end{array}} \right].
\end{equation}
\end{subequations}
where $a'$ and $b'$ are the lattice constant along the $x$ and the $y$ directions, respectively.

	\begin{table}
		\begin{center}
			\caption{Parameter values.}
			\label{tab1}
			\begin{tabular}{||c | c | c | c | c ||} 
				\hline
				Parameter & Na$_2$IrO$_3$ & $\alpha$-RuCl$_3$ \\
				\hline\hline
				$U$ (eV) & 1.7 & 3\\
				\hline
				$J_H$ (eV) & 0.3 & 0.6\\
				\hline
				$\lambda_{SOC}$ (eV) & 0.4 & 0.15\\
				\hline
				$\gamma_1$ (meV) & -22.9 & -19.8\\
				\hline
				$\gamma_2$ (meV) & -27.6 & -17.5\\
				\hline
				$\gamma_3$ (meV) & -27.2 & -12.5\\
				\hline
				$t_1 (meV)$ & 33.1 & 50.9\\
				\hline
				$t_2$ (meV) & 264.3 & 158.2\\
				\hline
				$t_3$ (meV) & 26.6 & -154\\
				\hline
				$t_4$ (meV) & -11.8 & -20.2\\
				\hline
				$t_5$ (meV) & -19.4 & -103.1\\
				\hline
				$t_6$ (meV) & 29.9 & 44.9\\
				\hline
				$t_7$ (meV) & 47.6 & 45.8\\
				\hline
				$t_{8}$ (meV) & -21.4 & -15.1\\
				\hline
				$t_{9}$ (meV) & -25.4 & -10.9\\
				\hline
				$t_{10}$ (meV) & 269.3 & 162.2\\
				\hline
			\end{tabular}
		\end{center}
		\begin{flushleft}
			$^\dag${\footnotesize The parameter values are taken from Ref. \cite{PhysRevB.93.214431}.}
		\end{flushleft}
	\end{table}

\section{NEGF Formalism}
\label{AppNEGF}
This section discusses the non-equilibrium Green's function (NEGF) method \cite{Datta1995} used for quantum-transport calculations in this paper.

\subsection{Self-Energy of Contacts}
The self-energy matrices $\Sigma_{1,2}$ of the left and right contacts on an arbitrary two-dimensional lattice is calculated using the surface Green's function $g_{1,2}$, respectively. $g_{1,2}$ are calculated by iteratively solving the following equations:
\begin{subequations}
\begin{equation}
    \left[g_1\right]=\left(E+j0^+\right)I-\left[\alpha\right]-\left[\beta\right]^\dag \left[g_1\right] \left[\beta\right],
\end{equation}
and
\begin{equation}
    \left[g_2\right]=\left(E+j0^+\right)I-\left[\alpha\right]-\left[\beta\right] \left[g_2\right] \left[\beta\right]^\dag.
\end{equation}
\end{subequations}
where $\left[ \alpha  \right]$ and $\left[ {{\beta }} \right]$ are from the following form of the Hamiltonian
\begin{equation}
    \begin{array}{l}
\mathcal{H}= \left[ \alpha  \right] + \left[ {{\beta }} \right]{e^{jk}} + \left[ {\beta } \right]^\dag{e^{ - jk}},
\end{array}
\end{equation}

Using the solutions, $g_{1,2}$, of the above equations, the self-energy functions are calculated as
\begin{subequations}
\begin{equation}
\label{self_energ1}
    \Sigma_1=\left[\beta\right] \left[g_1\right] \left[\beta\right]^\dag,
\end{equation}
and
\begin{equation}
\label{self_energ2}
    \Sigma_2=\left[\beta\right]^\dag \left[g_2\right] \left[\beta\right].
\end{equation}
\end{subequations}

\subsection{NEGF Quantities}
We calculate the following quantities:

\begin{itemize}
	\item Green's function:
	\begin{equation}
	\mathcal{G}=\left[E\mathcal{I}-\mathcal{H}-\Sigma\right]^{-1},
	\end{equation}
	with $\Sigma=\Sigma_1+\Sigma_2+\Sigma_0$. $\mathcal{H}$ is the Hamiltonian of the system as discussed earlier and $\mathcal{I}$ is an identity matrix. $\Sigma_0$ is the self-energy function that self-consistently takes into account the dephasing in the system (see Appendix \ref{AppNEGF}.\ref{subsec_dephas}).
	
	\item Spectral function:
	\begin{equation}
	\label{spec_func}
	\mathcal{A}=\mathcal{G}\,\Gamma\, \mathcal{G}^\dagger,
	\end{equation}
	with $\Gamma=\Gamma_1+\Gamma_2$ and $\Gamma_{1,2}$ are broadening functions which represent the anti-Hermitian part of $\Sigma_{1,2}$ i.e. $\Gamma_{1,2}=j\left(\Sigma_{1,2}-\Sigma_{1,2}^\dagger\right)$.
	
	\item Correlation function:
	\begin{equation}
	\label{e_func}
	G^n=\mathcal{G} \Sigma^{in} \mathcal{G}^\dagger,
	\end{equation}
	with $\Sigma^{in}=\Sigma^{in}_1+\Sigma^{in}_2+\Sigma^{in}_0$ being the in-scattering function. $\Sigma^{in}_0$ is the in-scattering function that self-consistently takes into account the dephasing in the system (see Appendix \ref{AppNEGF}.\ref{subsec_dephas}).
	
	\item In-scattering function:
	\begin{equation}
	\Sigma^{in}_{1,2}=\Gamma_{1,2}f_{1,2}.
	\end{equation}
	with $f_{1,2}$ being the Fermi occupation factors of contacts 1 and 2, given by
	\begin{subequations}
	\label{f_func}
	\begin{equation}
	f_{1}=\dfrac{1}{1+\exp\left({\dfrac{E-\mu_{1}+V_0/2}{k_BT}}\right)},
	\end{equation}
	and
	\begin{equation}
	f_{2}=\dfrac{1}{1+\exp\left({\dfrac{E-\mu_{1}-V_0/2}{k_BT}}\right)}.
	\end{equation}
	\end{subequations}
	Here, $\mu_{1,2}$ are the equilibrium electrochemical potentials of contacts 1 and 2, $V_0$ is the applied voltage between the two contacts, $k_B$ is the Boltzmann constant, and $T$ is the temperature.
\end{itemize}

\subsection{Dephasing}
\label{subsec_dephas}

We have included isotropic momentum and phase relaxation due to the static Coulomb interaction in the system. The self-energy and in-scattering functions that self-consistently take into account the dephasing are given by
\begin{subequations}
	\begin{equation}
	\Sigma_0=\mathcal{W}\,\mathcal{G},
	\end{equation}
	\begin{equation}
	\text{and, } \Sigma_0^{in}=\mathcal{W}\,G^n,
	\end{equation}
\end{subequations}
where $\mathcal{W}_{m,n}=\langle\,U_m\,U_n^*\rangle$ is calculated from the random potential at $m^\text{th}$ and $n^\text{th}$ lattices.

\section{Minimal model for $\text{Na}_2\text{IrO}_3$ and $\alpha\text{-RuCl}_3$}
\label{MinimalMOdel}

\begin{figure}
	%\vspace*{1cm}
	\includegraphics[width=0.4 \textwidth]{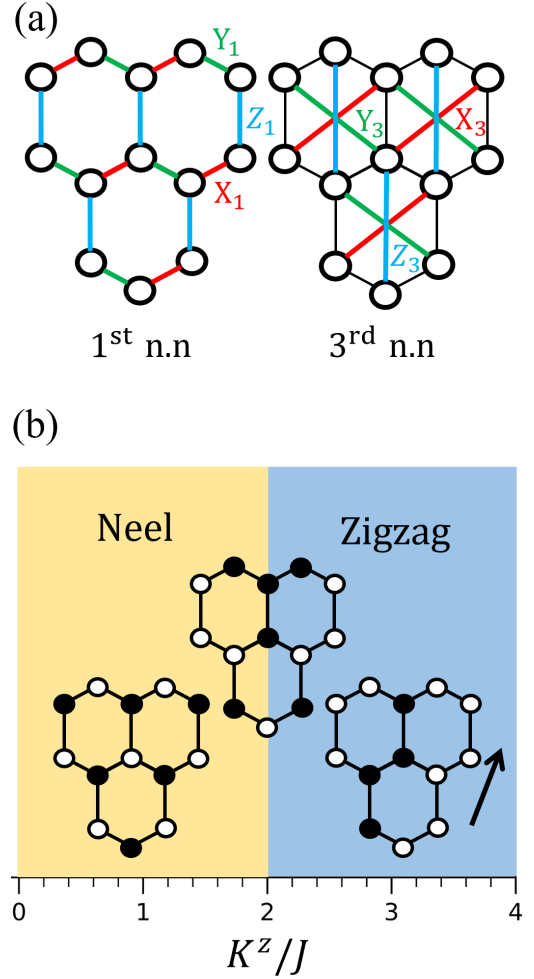}
	\centering
	\caption{(a) Indicating all the first and third nearest neighbor interactions. The interactions are color-coded according to the Kitaev interactions. (b) Ground state spin configurations for the minimal model Hamiltonian of $\text{Na}_2\text{IrO}_3$ when $K/J$ is swept from 0 to 3. Filled circle represents down spin and empty circle represents up spin . The spin configurations are used as an input for the NEGF transport model.}\label{Suppfig1}
\end{figure}

In this section, we discuss the Heisenberg-Kitaev Hamiltonian formulation for $\text{Na}_2\text{IrO}_3$ and $\alpha\text{-RuCl}_3$. The minimal model for $\text{Na}_2\text{IrO}_3$ is given by \cite{PhysRevB.93.214431}

\begin{equation}
    H = \sum_{\text{1st nn}} \left(J \mathbf{S_i} \cdot \mathbf{S_j} + K S_i^\gamma S_j^\gamma  \right)+ \sum_{\text{3rd nn}} J' \mathbf{S_i} \cdot \mathbf{S_j}
\end{equation}

The interactions for the above Hamiltonian can be visualised in Fig. \ref{Suppfig1}. Various level of theories including exact diagonalisation and perturbation theory were implemented to extract the Heisenberg and Kitaev parameters from the tight-binding electronic Hamiltonian\cite{PhysRevB.93.214431}. The parameters indicate that in case of $\text{Na}_2\text{IrO}_3$, the first nearest neighbor ferromagnetic Kitaev interaction term $K$ dominates the first nearest neighbor antiferromagnetic Heisenberg term. Meanwhile, the second nearest neighbor Heisenberg and Kitaev interaction terms are suppressed due to the interference of various second and third order hopping parameter terms \cite{PhysRevB.93.214431}. The values used to simulate $\text{Na}_2\text{IrO}_3$ are $(J', K, J) \sim (0.4, -1, 0.04)$ meV. We solve a 13 spin lattice system in a honeycomb structure containing 3 hexagons as shown in Fig. \ref{Suppfig1}. Ground state spin lattice configurations are obtained by sweeping the Hamiltonian parameter $K/J$ from 0 to 3 and exactly diagonalising the full Hamiltonian matrix. We observe for $K/J < 2$ that the lowest energy spin configuration is a N\'eel type antiferromagnet while for $K/J > 2$ it is a zigzag type antiferromagnet. At $K/J = 2$ interface, the ground state is a superposition of the N\'eel type and the zigzag type antiferromagnets, as shown in Fig. \ref{Suppfig1}. At $\text{Na}_2\text{IrO}_3$ Hamiltonian parameter values, the ground state spin configuration is found to be zigzag type antiferromagnet.

\section{Conditions for chiral edge conduction due to an anisotropic interaction}
\label{AppA}

This section elaborates the 1D analytical arguments that shows that the topological edge state formation for $K\geq2J$ is an intrinsic property of the Eq. \eqref{Full_KH_Model} in the main manuscript.

\subsection{Derivation of Eq. \eqref{KH_1D_JW} in the main manuscript}

We expand Eq. \eqref{KH_Model_1D} in the main manuscript as
\begin{equation}
\begin{aligned}
\label{1dchain1}
    \mathcal{H} = \left( {J + \frac{{{K^x}}}{2}} \right)\sum\limits_m {S_{m + 1}^xS_m^x}  + \left( {J + \frac{{{K^y}}}{2}} \right)\sum\limits_m {S_{m + 1}^yS_m^y}  \\+ J\sum\limits_m {S_{m + 1}^zS_m^z}  - \frac{{{K^x}}}{2}\sum\limits_m {{{\left( { - 1} \right)}^m}S_{m + 1}^xS_m^x}  \\+ \frac{{{K^y}}}{2}\sum\limits_m {{{\left( { - 1} \right)}^m}S_{m + 1}^yS_m^y}.
\end{aligned}
\end{equation}

The spin-raising and spin-lowering operators are given by
\begin{subequations}
\begin{equation}
    \;\;\;\;\;\;\;\;\;{S^ + } = {S^x} + i{S^y},
\end{equation}
\begin{equation}
    \text{and,}\;\;\;{S^ - } = {S^x} - i{S^y},
\end{equation}
\end{subequations}
respectively, which makes Eq. \eqref{1dchain1} as
\begin{equation}
\label{chain2}
\begin{array}{l}
{\cal H} = \dfrac{{{K^x} - {K^y}}}{8}\sum\limits_m {\left( {S_{m + 1}^ + S_m^ +  + S_{m + 1}^ - S_m^ - } \right)}  + J\sum\limits_m {S_{m + 1}^zS_m^z} \\
 + \dfrac{1}{2}\left( {J + \dfrac{{{K^x} + {K^y}}}{4}} \right)\sum\limits_m {\left( {S_{m + 1}^ + S_m^ -  + S_{m + 1}^ - S_m^ + } \right)} \\
 - \left( {\dfrac{{{K^x} + {K^y}}}{8}} \right)\sum\limits_m {{{\left( { - 1} \right)}^m}\left( {S_{m + 1}^ + S_m^ +  + S_{m + 1}^ - S_m^ - } \right)} \\
 - \left( {\dfrac{{{K^x} - {K^y}}}{8}} \right)\sum\limits_m {{{\left( { - 1} \right)}^m}\left( {S_{m + 1}^ + S_m^ -  + S_{m + 1}^ - S_m^ + } \right)} .
\end{array}
\end{equation}
Note that Eq. \eqref{chain2} consists of a Heisenberg term, an Ising term, and a sign-altering double-spin-flip functuation term \cite{Agrapidis2018,PhysRevB.90.035113}.

In 1D, $S=\frac{1}{2}$ chains can be represented as fermions with $\left| \uparrow \right\rangle\equiv f^\dagger \left| \downarrow \right\rangle$ and $\left| \downarrow \right\rangle\equiv f \left| \uparrow \right\rangle$, and using the Jordan-Wigner transformation \cite{PhysRevB.80.125415, PhysRevB.101.094403}, $S^+$, $S^-$, and $S_z$ can be represented as
\begin{subequations}
\begin{equation}
    {S^ +_m } = {f^\dag_m } e^{i\pi\sum_{l<m} n_m},
\end{equation}
\begin{equation}
    {S^ -_m } = {f_m } e^{-i\pi\sum_{l<m} n_m},
\end{equation}
\begin{equation}
    \text{and, }\;\;\;{S^z_m} = {f^\dag_m }f_m - \frac{1}{2},
\end{equation}
\end{subequations}
using which we can write 
\begin{subequations}
\label{some_relations1}
    \begin{equation}
        \sum\limits_m {\left( {S_{m + 1}^ + S_m^ +  + S_{m + 1}^ - S_m^ - } \right)}  = \sum\limits_m {\left( {f_{m + 1}^\dag f_m^\dag  + {f_m}{f_{m + 1}}} \right)},
    \end{equation}
    \begin{equation}
        \sum\limits_m {\left( {S_{m + 1}^ + S_m^ -  + S_{m + 1}^ - S_m^ + } \right) = \sum\limits_m {\left( {f_{m + 1}^\dag {f_m} + f_m^\dag {f_{m + 1}}} \right)} },
    \end{equation}
    \begin{equation}
        \text{and,}\;\;\;\sum\limits_m {S_{m + 1}^zS_m^z} = \sum\limits_m {{n_{m + 1}}{n_m}}  - 2\sum\limits_m {{n_m}}  + \frac{N}{4},
    \end{equation}
\end{subequations}
where $n_m=f^\dagger_m f_m$ and $N$ is the number of 1D lattice points. Note that we have ignored the constant term $N/4$. We apply \eqref{some_relations1} in Eq. \eqref{chain2} to get Eq. \eqref{KH_1D_JW} in the main manuscript.

\subsection{Derivation of Eq. \eqref{KH_1D_kspace} in the main manuscript}
We use the Fourier transformation
\begin{equation}
    {f_m} = \frac{1}{{\sqrt N }}\sum\limits_{k'} {{s_{k'}}{e^{i{k'}{m}}}},
\end{equation}
where $N$ is the number of lattice points in the atomic chain, $k'=k-k_0$ with $k$ being the wavevector and $k_0$ being the wavevector at the band minimum. 

We transform Eq. \eqref{KH_1D_JW} from real space into the momentum space. The transformation yields
\begin{subequations}
\label{some_relations2}
\begin{equation}
\begin{aligned}
    \sum\limits_m \left( f_{m + 1}^\dag f_m^\dag  + f_m f_{m + 1} \right)  = \sum \limits_{k'} s_{ - k'}^\dag s_{k'}^\dag e^{ik'} \\+ \sum \limits_{k'} s_{k'} s_{-k'} e^{ - ik},
\end{aligned}
\end{equation}
\begin{equation}
\begin{aligned}
    \sum\limits_m \left( f_{m + 1}^\dag f_m^\dag  + f_m f_{m + 1} \right)  = \sum\limits_{k'} s_{-k'}^\dag s_{k'}^\dag \left( e^{ik'} - e^{-ik'} \right) \\+ \sum \limits_{k'} s_{k'} s_{-k'}\left( e^{ - ik'} - e^{ik'} \right),
\end{aligned}
\end{equation}
\begin{equation}
\begin{aligned}
    \sum\limits_m \left( { - 1} \right)^m &\left( f_{m + 1}^\dag f_m^\dag  + f_m f_{m + 1} \right)  \\ &=  - \sum\limits_{k'} \left( s_{ - k'}^\dag s_{k'}^\dag e^{i k'} + s_{k'} s_{ - k'}e^{ - ik'} \right),
\end{aligned}
\end{equation}
\begin{equation}
\begin{aligned}
    &\sum\limits_m {{{\left( { - 1} \right)}^m}\left( {f_{m + 1}^\dag f_m^\dag  + {f_m}{f_{m + 1}}} \right)}  \\&=  - {\textstyle{1 \over 2}}\sum\limits_{k'} {s_{ - {k'}}^\dag s_{k'}^\dag \left( {{e^{i{k'}}} - {e^{ - i{k'}}}} \right)}  \\&- {\textstyle{1 \over 2}}\sum\limits_{k'} {{s_{k'}}{s_{ - {k'}}}\left( {{e^{ - i{k'}}} - {e^{i{k'}}}} \right)},
\end{aligned}
\end{equation}
\begin{equation}
\begin{aligned}
    \sum\limits_m {\left( {f_{m + 1}^\dag {f_m} + f_m^\dag {f_{m + 1}}} \right)}  = \sum\limits_{k'} {s_{k'}^\dag {s_{k'}}\left( {{e^{i{k'}}} + {e^{ - i{k'}}}} \right)},
\end{aligned}
\end{equation}
\begin{equation}
\begin{aligned}
    \sum\limits_m {{{\left( { - 1} \right)}^m}\left( {f_{m + 1}^\dag {f_m} + f_m^\dag {f_{m + 1}}} \right)}  \\=  - \sum\limits_{k'} {s_{k'}^\dag {s_{k'}}\left( {{e^{i{k'}}} + {e^{ - i{k'}}}} \right)},
\end{aligned}
\end{equation}
\begin{equation}
\sum\limits_m {{n_m}}  = \sum\limits_{k'} {s_{k'}^\dag {s_{k'}}},
\end{equation}
\begin{equation}
\label{interaction_J}
    \text{and,}\;\;\;\sum\limits_m {n_{m+1}}{{n_m}}  = \frac{1}{N}\sum\limits_{k',q} {J {{e^{-iq}}} s_{k'-q}^\dag {s_{k'}^\dag} {s_{k'-q}}}s_{k'}.
\end{equation}
\end{subequations}
where $q$ is the wavevector. We apply Eqs. \eqref{some_relations2} in Eq. \eqref{KH_1D_JW} to get Eq. \eqref{KH_1D_kspace} in the main manuscript.

\subsection{Energy gap}
We find the dispersion relation of Eq. \eqref{TB_KH_matrix} by solving $|\mathrm{H}-E\mathcal{I}|=0$, which is given by
\begin{equation}
\label{egap1}
    E=\pm\sqrt{\left(J-\dfrac{2J+K}{4}\cos k'\right)^2+\left(\dfrac{K}{4}\sin k'\right)^2}.
\end{equation}

Near the band minimum, i.e., $k\rightarrow k_0$, we can the write the following expression write from Eq. \eqref{egap1}.
\begin{equation}
    E=\pm\dfrac{2J-K}{4}.
\end{equation}

Thus, the energy gap is given by
\begin{equation}
    E_{gap}=\pm\dfrac{2J-K}{2}.
\end{equation}

\bibliography{main}

\end{document}